\documentclass[namedreferences]{kluwer}

\usepackage{epsfig}
\usepackage{doc}

\begin{document}
\begin{article}

\begin{opening}

\title{The robust detection of stars on CCD images}

   \author{Filip \surname{Hroch}\email{hroch@physics.muni.cz}}
   \runningauthor{Filip Hroch}
   \runningtitle{The robust detection of stars on CCD images}

   \institute{Astronomical Institute,  Faculty of Science, 
              Masaryk University,\\ Kotl\'{a}\v{r}sk\'{a} 2, 
              611~37 Brno, Czech Republic}


\begin{abstract}
Two parameters are developed to analyze the CCD images from 
ground-based and/or space telescopes. The first parameter,
deduced from the intensity profile of the object $sharp$, is 
useful to resolve stars and hot pixels. The second parameter
$shape$ distinguishes the stars from the background cosmic-ray 
events using geometric characteristics defined by its shapes. 
The parameters are applied to a simulated OMC/INTEGRAL
image and a HST image.
\end{abstract}


\end{opening}


\section{Introduction}
The general problem of the fully automatic star detecting 
algorithms on the CCD images is the successful recognition 
of real star images or star-like sources in general from other 
types of objects and defects. These false objects originate 
from non-ideal characteristics
of the CCD instrument (noise, bad column and hot pixels)
or from the environment (nebulae, galaxy or cosmic ray events
known as 'cosmics'). The frequent 
requirement is to find real star images on a diffuse
background or in a crowded star field.
Therefore, the detecting algorithm cannot use a simple intensity
comparing principle, because the intensity of non-stellar objects can 
be comparable to star intensities. 

A number of different methods for the automatic star detection 
on astronomical images has been used during the last two decades. For example, 
the method based on wavelet transforms applied to satellite 
ROSAT image processing was published by Damiani~et~al.~\cite{Damiani}. 
The automatic  search and classification were implemented in FOCAS 
routine by Valdes~et~al.~\cite{Valdes}. The PC~Vista by Trefers and 
Richmod~\cite{Treffers} is a package for photometry with
a simple algorithm for star detection. The well-known program
DAOPHOT by Stetson~\cite{Stetson} contains two parameters
$sharp$ and $round$ and a highly robust algorithm for star detection.
The determination of these parameters originates from specific
characteristics of the sources on CCD images, but both
parameters are not sensitive enough to resolve stars and cosmics
on the CCD images exposed in focus of a space telescope
and/or camera such as the OMC experiment on-board INTEGRAL 
(Hermsen and Winkler~\cite{integral}). 


\section{Distinguishing hot pixels}

The large part of the information describing the detected object 
is included in its intensity profile, but the profile varies 
widely with weather, telescope focus, etc. changes. The intensity
distribution cannot be described by an analytical
function, because it depends on many non-controllable
factors. Fortunately, the real measured profiles are (in the first
approximation) close to the two-dimensional Gaussian
profile both for the ground-based (Moffat~\cite{Moffat}) and 
space telescopes.
 
\begin{figure}[p]
 \centerline{\epsfig{file=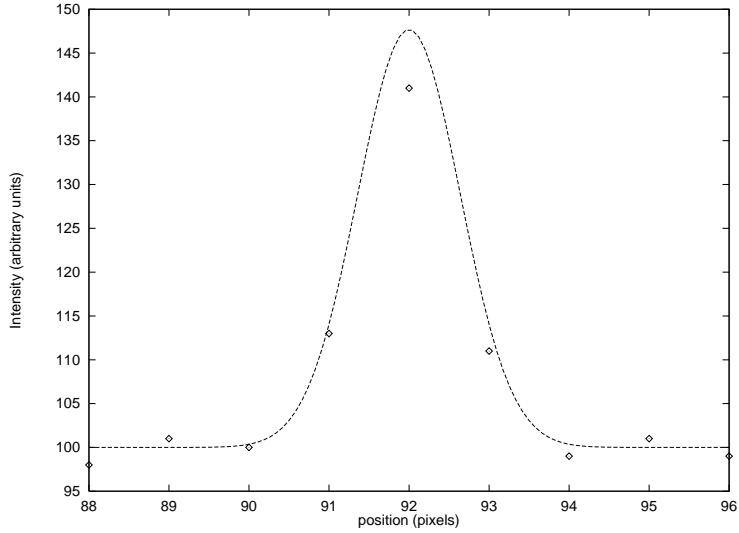,width=10cm}}

 \caption[]{The intensity profile for typical star image. 
           Intensity in arbitrary units, $x$ coordinate in pixels. 
           The Gaussian best fit is plotted with parameters: $G_0 = 50.1$, 
           $h = 0.64$, $x_c = 92.0$, $y_c = 76.8$.
           The background intensity is $100$ and the observed intensity 
           peak $I_0 = 41$ satisfies $G_0 > I_0$.  }
 \label{fig1}
\end{figure}

\begin{figure}[p]
 \centerline{\epsfig{file=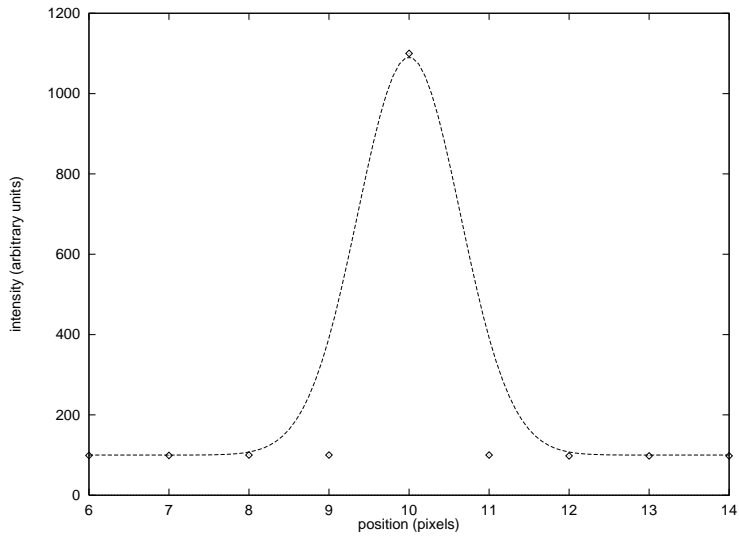,width=10cm}}

 \caption[]{The intensity profile for a hot pixel. Intensity in arbitrary
           units, $x$ coordinate in pixels. The Gaussian best fit
           is plotted 
           with parameters: $G_0 = 991.3$, $h = 0.64$, $x_c = 10.0$,
           $y_c = 90.0$.
           The background intensity is $100$ and the observed intensity 
           peak $I_0 = 1000$ satisfies $G_0 < I_0$.  }
 \label{fig2}
\end{figure}

From this point of view, we can easy distinguish between the hot 
pixel (a~single pixel with very different intensity compared
to surrounding pixels) and the real star. The star image with the Gaussian 
profile is usually spread over several pixels. For illustration, the 
typical 1-dimensional profiles of the star and of the hot pixel are shown 
in  Figures~\ref{fig1} and \ref{fig2}. The maximum intensity
of the object  $I_{i_0 j_0}$ is at the pixel $(i_0, j_0)$. 
The center of the star is at position
\begin{equation} \label{e1}
  x_c = \frac{\sum I_{ij} \cdot i}{\sum I_{ij}} \qquad  
  y_c = \frac{\sum I_{ij} \cdot j}{\sum I_{ij}}, 
\end{equation}
where summations are evaluated over a small area around 
the central pixel $(i_0, j_0)$ --- for example over $5 \times 5$ pixels. 
This notation is also used below.
Approximation of the observed intensity by the two-dimensional 
Gaussian profile is the next step of our algorithm  
\begin{equation}
  G(i,j) = G_0 e^{-[(i - x_c)^2 + (j - y_c)^2]/2 h^2} + \, b
\end{equation}
The parameter $G_0$ is the estimated maximum of the star profile,
$b$ is the local sky background. The parameter $h$ is the
'width' of the typical star and its numerical value must be
known {\em a priori}. The stars are axis-symmetrical objects 
according to this formula. Note, that the $G_0$ parameter
is equivalent to, but not quite identical with the Stetson's 
parameter $H_{i_0 j_0}$ (integer values are used for $x_c, y_c$,
i.e. $x_c \equiv i_0, y_c \equiv j_0$).
If we suppose that the statistical distribution of the observed
intensity has a normal distribution, we can use the least
square method
\begin{equation} \label{e5}
  S = \sum (G(i,j) - I_{i,j})^2 \; \rightarrow \; \rm min
\end{equation}
for the estimation of the unknown $G_0$ and (further unused) $b$.

Further, we can estimate the central intensity
$I_0$ from the maximum intensity $I_{i_0 j_0}$ by subtracting
the local background intensity 
$B_{i_0 j_0}$ from $I_{i_0 j_0}$:
\begin{equation} \label{e2}
  I_0 = I_{i_0 j_0} - B_{i_0 j_0}
\end{equation}
where
\begin{displaymath} 
  B_{i_0 j_0} = \frac{\sum\limits_{(i,j) \not = (i_0, j_0)}
                      w_{ij} I_{ij}}{\sum\limits_{(i,j) 
                      \not = (i_0, j_0)} w_{ij}} 
\end{displaymath}
($w_{ij}$ are arbitrary weights) 
and compare its numerical value in parameter $sharp$:
\begin{equation}
  sharp \equiv \frac{I_0}{G_0}.
\end{equation}
The Figures~\ref{fig1} and \ref{fig2} indicate the typical values
of this parameter. Its value is in agreement with the inequality
$G_0 > I_0$ for the star and with $G_0 < I_0$ for the hot pixel. 
We can hence deduce the range of the parameter $sharp$  as  
$0 < sharp < 1$ for the star. 


\section{Distinguishing cosmic-ray events}

The cosmic ray events frequency on the CCD images exposed at 
a ground-based observatory is low and, moreover, its profiles 
are narrower than the star profiles. The image processing 
algorithm can then easily resolve the images of the 
stars and those of the cosmic-ray events. But CCD images 
from satellites contain a higher number of the cosmic-ray
event tracks and the numerical value of the $sharp$ parameter, computed 
for the stars, is close to the numerical value for cosmics. The 
construction of the confidentiality test for its distinguishing with 
only the $sharp$ parameter is not possible. 

The star isophotes of the ideal images have the shape  of  
concentric circles, but the cosmics or the bad columns are 
represented in general by ellipses with the major semi-axis
at some angle to the $x$ coordinate.
The mathematical description of an object in image is derived
from this characteristics.

The most effective way is the description of the object by its 
2nd moments
\begin{equation} \label{e3}
   h_x^2 = \frac{\sum I_{ij} (i - x_c)^2}{\sum I_{ij}}, \qquad
   h_y^2 = \frac{\sum I_{ij} (j - y_c)^2}{\sum I_{ij}}
\end{equation}
and
\begin{equation} \label{e4}
   h_{xy} = \frac{\sum I_{ij} (i - x_c)(j - y_c)}{\sum I_{ij}}
\end{equation}
where the summations  are evaluated over the same small area around
$(i_0, j_0)$. Now, we can compute the lengths of the object 
semi-axes as the 'eigenvalues' problem. Its lengths satisfy 
the equation
\begin{displaymath}
  \left| 
  \begin{array}{lr}
    \lambda - h_x & h_{xy} \\
    h_{xy} &  \lambda - h_y
  \end{array}
  \right| = 0
\end{displaymath}
Resolving of this equations for $\lambda$ gives the sizes of 
semi-axes as
\begin{displaymath}
 \lambda_{\pm} = \frac{1}{2} 
   \left[ (h_x + h_y) \pm \sqrt{(h_x - h_y)^2 + 4 h_{xy}^2} \right]
\end{displaymath}
The numerical values of its parameters (lengths of semi-axes)
are different for elongated object, but approximately the same
for stars. With respect to their character we define the parameter 
\begin{equation}
  shape \equiv 2 \frac{\sqrt{(h_x - h_y)^2 + 4 h_{xy}^2} }{ h_x + h_y}
\end{equation}
The limits of this parameter are:
\begin{itemize}
\item $h_x \rightarrow h_y, h_{xy} \rightarrow 0 :
           shape \rightarrow 0$, the star case
\item $h_x \leftrightarrow h_y, h_{xy} \not = 0 : 
      shape \rightarrow 4 h_{xy} / (h_x + h_y)$, the cosmics case
\item $h_x \not = h_y, h_{xy} \rightarrow 0 : |shape| \rightarrow
      2 |(h_x - h_y)/(h_x + h_y)| \gg 0$, the bad column case,
      $shape$ parameter is then equivalent to the $round$ parameter 
      of Stetson~(1987)
\end{itemize}

The parameter $shape$ describes the shape of the object. The circular 
objects are interpreted this way as stars and the very elongated object as  
cosmic-ray events or bad columns. Note, that the $shape$ parameter 
represents direct generalization of the $round$ parameter described 
in Stetson~(1987). The $round$ parameter is capable of
measuring elongation only parallel to rows or columns of the image.
It was intended to identify bad rows or columns, not cosmics. 


\begin{figure}[p]
 \begin{center}
 \unitlength = 1.0 cm
 \begin{picture}(8,8)
 \framebox(8,8){ \psfig{figure=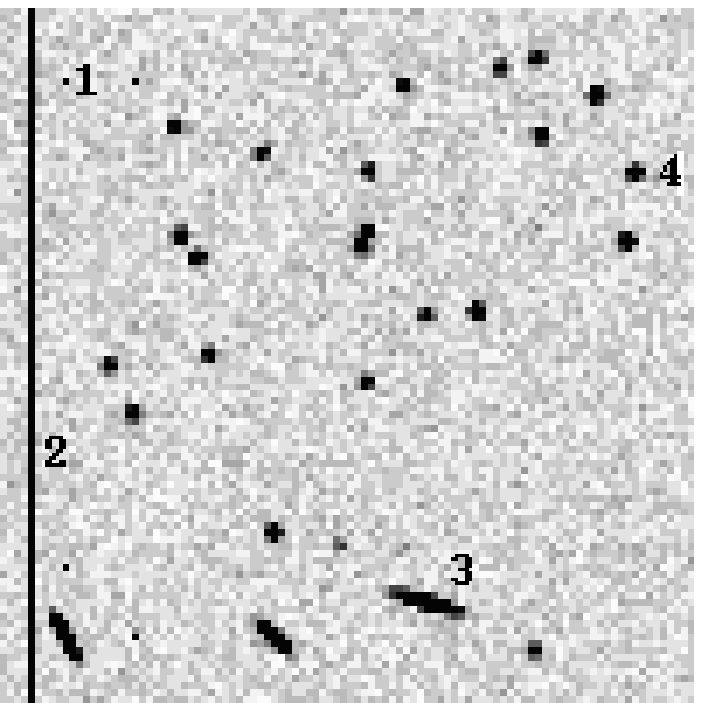,width=8cm} }
 \end{picture}
 \end{center}


 \caption[]{The simulated image with stars and other sources
           (legend: No. 1 --- hot pixel, No. 2 --- bad column,
                   No. 3 --- cosmics,    No. 4 --- star).}
 \label{fig3}
\end{figure}

\section{The simulated image}

In this section we describe the use of the parameters 
mentioned in previous sections to analyze
the simulated image of the star field. The image 
(see Fig.~\ref{fig3}) is generated as the simulation of the 
field of view of the OMC (Optical Monitoring Camera) prepared for
the ESA satellite INTEGRAL. 
The PSF function of the stars in the image is defined
as the two-dimensional Gaussian profile with 70\% of 
the light energy in the central pixel, according to the PSF
of the flight optics. This results in $h = 0.64$ for the stars 
in the simulated image, while their
centers are uniformly spread over the unit interval. There are also
some other sources in the image, namely the bad column, hot pixels
and a few cosmics (see Fig.~\ref{fig3} and
Table~\ref{tab}). The image is critically under-sampled.

The image left-bottom corner has coordinates $(0,0)$ and 
right-top one $(100,100)$ in pixels. The bad column is situated 
at $x_c = 5$. The hot pixels occupy positions $(10,90)$, $(10,20)$,
$(20,10)$, $(20,90)$ and the centers of the three cosmics appear at 
$(10,10)$, $(40,10)$, $(62,15)$. All other sources represent  
real stars. The intensity of the sky is $100$ with the Gaussian 
noise $1.1$ (all intensities in arbitrary units). Object No.~4 
(star) and object No.~1 (hot pixel) were used for construction
of the profiles shown in Fig.~\ref{fig1} and~\ref{fig2}.
 
The computed characteristics for the selected sources in the simulated 
image are listed in Table~\ref{tab}. The center of the object
is at coordinates $x_c, y_c$, see formulae (\ref{e1}), the intensity
peak at pixel $(i_0, j_0)$ is related to $I_0$ according to (\ref{e2}).
The moments computed from (\ref{e3}) and (\ref{e4}) are $h_x$, $h_y$,
$h_{xy}$ and the parameter $G_0$ is obtained by minimization of the 
sum (\ref{e5}). The last three columns contain the values of the
parameters $sharp, round$ and $shape$.

The values of the numerical differences of the $sharp$ parameter for star and 
hot pixel are small because $h$ is small, but the upper bound at 1 separates 
stars and hot pixels to obviously different groups. 
The values of the $round$ parameter demonstrate  small 
difference between cosmics and stars. On the contrary the values of 
the $shape$ parameter of the stars (or hot pixels) and of the cosmics 
as well as that of the bad columns are clearly incompatible.

\begin{table}[t]
\caption[]{The parameters of objects in the simulated image
           (legend: No. 1 --- hot pixel, No. 2 --- bad column,
                   No. 3 --- cosmics,    No. 4 --- star).}
  \label{tab}
\begin{tabular}{rrrrrrrrrrrr}
\hline
No. & $x_c$ &  $y_c$  &  $I_0$  &  $h_x$  &  
$h_y$  & $h_{xy}$ & $G_0$ & $round$ & $sharp$  & $shape$ \\
\hline
1 &  10.0 & 90.0 & 1000 & 0.17 & 0.17 & 0.00 & 991.4 & 
  0.02 & 1.01 & 0.06 \\
2 & 5.0 &  5.0 & 1000 & 0.07 & 1.41 & 0.00 & 865.3 & 
  $-1.81$ & 0.91 & 1.80 \\
3 &  62.0 & 15.0 &  199 & 1.30 & 0.62 & $-0.46$ & 194.2 &
  0.71 & 0.92 & 1.20 \\
4 &  92.0 & 76.8 &  41 &   0.69 &   0.66 &   0.03  &  50.1 & 
 0.04 & 0.80 & 0.10  \\
\hline
\end{tabular}
\end{table}


\begin{figure}[p]
 \begin{center}
 \unitlength = 1.0 cm
 \begin{picture}(8,8)
 \framebox(8,8.05){ \psfig{figure=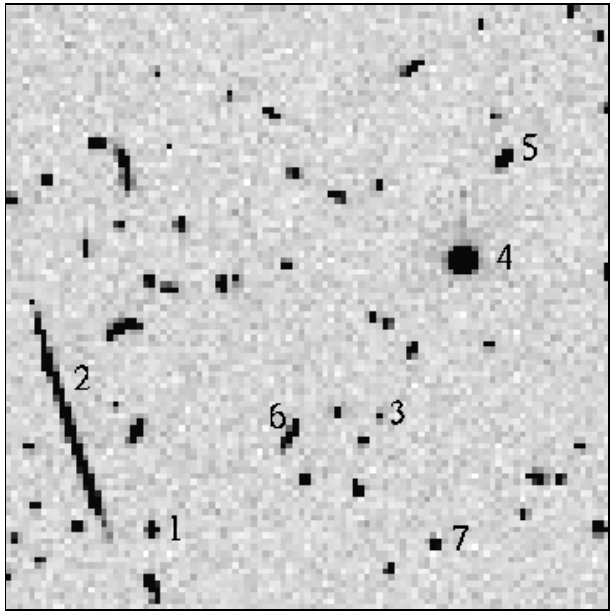,width=8cm} }
 \end{picture}
 \end{center}


 \caption[]{The HST image with stars and cosmics
           (legend: No.~1 --- star, No.~2 ---cosmics, No.~3 --- star,
                   No.~4 --- nebulosity(?), No.~5 --- double star,
                   No.~6 --- cosmics, No.~7 --- star).}

 \label{fig4}
\end{figure}


\section{The HST image}

A wide variety of different kinds of objects appears in 
the simulated image, but the real images can also contain others 
unpredictable defects.
We test the applicability of the parameters on an example of a real image 
(Fig.~\ref{fig4}). We have extracted and zoomed 
a small sub-window
of the image from HST. The image has been corrected on photometric corrections
(bias, flat field, etc.) and exhibits no hot pixels or bad 
columns. We set the value of the $h$ parameter to $h = 0.4$.

The list of derived characteristics for a selected objects
in this image is given in Table~\ref{tab2}. The meaning of the columns is 
identical with Table~\ref{tab}. The objects numbers correspond with 
Fig.~\ref{fig4}.
The intensity of the sky is $8.3$ with the Gaussian 
noise $1.5$ (all in arbitrary units).

Identifying by 'a visual method' leads to classify the objects 
No.~1, 3 and 7
as stars, No.~2 and 6 as cosmics, No.~4 as nebulosity and 
No.~5 is probably a double star. 
The values of the $sharp$ parameters are in the range valid for
stars (and hence indicating the absence of
hot pixels). 'The visual method' classification of stars and cosmics
confirms the values of the $shape$ parameter --- values for cosmics 
significantly exceeds 1. The $round$ parameter has little use 
for successful recognition 
between stars and cosmics. This example can be described as a typical one, 
because the stars, galaxies
and cosmics represent the most frequent objects in the HST images.
The double star case (No.~5) complicates this test and it can 
easily result in the failure of the whole procedure
--- the $shape$ parameter can be near the limit set in the algorithm
and we have to set it carefully.


\begin{table}[t]
\caption[]{The parameters of objects in the HST image
           (legend: No.~1 --- star, No.~2 ---cosmics, No.~3 --- star,
                   No.~4 --- nebulosity(?), No.~5 --- double star,
                   No.~6 --- cosmics, No.~7 --- star).}
  \label{tab2}
\begin{tabular}{rrrrrrrrrrrr}
\hline
No. & $x_c$ &  $y_c$  &  $I_0$  &  $h_x$  &  
$h_y$  & $h_{xy}$ & $G_0$ & $round$ & $sharp$  & $shape$ \\
\hline
1 & 25.0 & 13.8 & 406.00  &  0.33  &  0.54  &  0.00 &  774.39 & 
   -0.49 & 0.51 &  0.49  \\
2 &  9.0 & 38.0 & 724.00  &  0.41  &  0.95  & -0.26 &  717.75 &
   -0.79 & 0.96 &  1.11  \\
3 & 63.1 & 33.0 &  90.00  &  0.60  &  0.60  & -0.01 &  116.62 &
    0.00 & 0.70 & 0.04   \\
4 & 77.1 & 59.1 & 419.00  &  0.97  &  0.98  &  0.02 &  502.82 &
   -0.01 & 0.77 & 0.04   \\
5 & 83.8 & 75.9 & 450.00  &  0.63  &  0.76  &  0.31 &  723.16 &
   -0.19 & 0.59 & 0.92   \\
6 & 48.3 & 30.2 & 124.00  &  0.74  &  1.24  &  0.72 &  877.73 &
   -0.51 & 0.12 & 1.54   \\
7 & 72.3 & 11.9 & 307.00  &  0.53  &  0.38  &  0.00 & 1447.14 &
    0.32 & 0.20 & 0.32   \\
\hline
\end{tabular}
\end{table}


\section{Conclusions}

The two parameters $sharp$ and $shape$ described in this paper
are widely applicable to the robust detection of the star-like 
sources in images provided by array detectors. The primary reason 
for deriving the parameters was the development
of a reliable automatic algorithm for the recognition of stars
on CCD images. We can choose following criterion for this purpose it
the intensity of the central pixel of the examined object is above 
an arbitrarily defined 'level of significance' (threshold)
and the object lying far from the edge of the frame. Then
\begin{itemize}
\item [(a)] the object is a star for: $0 < sharp <1$ and $ 0 < shape < 1$.
\item [(b)] the object is a cosmics for: $ shape > 1$.
\item [(c)] the object is a bad column for: $ shape > 1$.
\item [(d)] the object is a hot pixel for: $ sharp > 1$.
\end{itemize}
This definition provides a strict mathematical criterion for detection 
and classification of the various kinds of objects on CCD images.

We might specify a range of the $sharp$ and $shape$ parameters 
for a star centered on a pixel as $0 < sharp <1$, $ 0 < shape < 1$.
The range of $sharp$ for a star centered on the edge between two 
pixels decrease to $0 < sharp <1/2$ and one for a star centered 
on the corner of four pixels decrease to $0 < sharp <1/4$, but 
the range for $shape$ increase to $ 0 < shape < 2$ in this both cases.

\begin{acknowledgements}
I am grateful to R. Hudec, T. Rezek and B. Lencov\'{a} for their 
correction of the manuscript and many helpful comments.     
This work has been supported by the Czech Ministry for 
Education and Youth, Project Nr.~ES~036 KONTAKT.
We have done many of computation at Supercomputing Center of Masaryk 
University in Brno. The HST image originates from 
the HST Archive at the Space Telescope Institute.

\end{acknowledgements}


\end{article}
\end{document}